\documentclass[twoside]{ilcws07}
\usepackage[latin1]{inputenc}
\usepackage[dvips]{graphicx,epsfig,color}
\usepackage{wrapfig,rotating}
\usepackage{amssymb,amsmath,array}

\begin{document}
\title{Discovery and Identification of Contactlike Interactions in
 Fermion-pair Production at ILC} %%
%***********************************************************************
% AUTHORS INFORMATION AREA
%***********************************************************************
\author{A. A. Pankov$^1$, N. Paver$^2$ and A. V. Tsytrinov$^1$
% Optional short acknowledgment: remove next line if non-needed
%\thanks{This is an optional funding source acknowledgment.}
% DO NOT MODIFY THE FOLLOWING '\vspace' ARGUMENT
\vspace{.3cm}\\
% Addresses and institutions (remove "1- " in case of a single institution)
1- ICTP Affiliated Centre, Pavel Sukhoi Technical University - Dept of
Physics\\
Gomel 246746 - Belarus
%% Remove the next three lines in case of a single institution
\vspace{.1cm}\\
2- University of Trieste and INFN-Trieste Section - Dept of Theoretical
Physics \\
34100 Trieste  - Italy\\
}
%%***********************************************************************
% END OF AUTHORS INFORMATION AREA
%*******************************************

\maketitle

\begin{abstract}
Non-standard scenarios described by effective contactlike
interactions can be revealed only by searching for deviations of
the measured observables from the Standard Model (SM) predictions.
If deviations were indeed observed within the experimental
uncertainty, the identification of their source among the
different non-standard interactions should be needed. We here
consider the example of the discrimination of gravity in
compactified extra dimensions (ADD model) against the four-fermion
contact interactions (CI). We present assessments of the
identification reach on this scenario, that could be obtained from
measurements of the differential cross sections for the fermionic
processes $e^+e^-\to{\bar f}f$, with $f=e,\mu,\tau,c,b$, at the
planned ILC.
\end{abstract}

\section{Non-standard effective interactions}

The non-standard contactlike local interactions we are going to
consider are all characterized by corresponding large mass scales
$\Lambda_{\alpha\beta}$ and $\Lambda_H$ to some inverse power that
specifically depends on the dimension of the relevant effective
local operators:
\par
{\bf a)} The compositeness inspired dim-6 four-fermion contact interactions
(CI):
\begin{equation}
{\cal L}^{\rm CI}=4\pi\sum_{\alpha,\beta}\hskip 3pt
\frac{\eta_{\alpha\beta}}{\Lambda^2_{\alpha\beta}} \left(\bar
e_\alpha\gamma_\mu e_\alpha\right) \left(\bar f_\beta\gamma^\mu
f_\beta\right),\qquad \eta_{\alpha\beta}=\pm1,0 ,
\label{CI}
\end{equation}
with $\alpha,\beta={\rm L,R}$ the helicities of the incoming and
outgoing fermions~\cite{Eichten:1983hw}. Generally, this kind of
models can describe exchanges between SM particles of very heavy
$W^\prime$, $Z^\prime$, leptoquarks, {\it etc}.
\par
{\bf b)} The ADD model of gravity in ``large'' compactified extra
dimensions~\cite{Arkani-Hamed:1998rs}, that can be parameterized
by the dim-8 contactlike interaction~\cite{Hewett:1998sn}:
\begin{equation}
{\cal L}^{\rm
ADD}=i\frac{4\lambda}{\Lambda_H^4}T^{\mu\nu}T_{\mu\nu},
\qquad \lambda=\pm 1.
\label{dim-8}
\end{equation}
Here, $T_{\mu\nu}$ is the energy-momentum of SM particles, and $\Lambda_H$
essentially represents a cut-off on the exchange (in 4 dimensions) of a
tower of Kaluza-Klein, spin-2, massive graviton excitations. For (sub)millimeter
extra dimensions, the mass $\Lambda_H$ scale may be expected to be of the TeV size.
\par
In principle, in addition to the Planck mass $M_D$ in $4+n$ dimensions, such
that $M_{\rm PL}=M_D^{1+n/2}R^{n/2}$ with $R$ the compactification radius,
there can exist one independent mass scale we denote generically as $\Lambda$, that represents the relative strength of tree {\it vs.} loop virtual graviton exchanges. In the naive dimensional approximation (NDA), the relation of
this extra scale to $\Lambda_H$ in Eq.~(\ref{dim-8}) is~\cite{Giudice:2003tu}:
\begin{equation}
\frac{1}{\Lambda_H^4}=\frac{\pi^{n/2}}{8\Gamma(n/2)}\hskip
3pt \frac{\Lambda_{\rm NDA}^{n-2}}{M_D^{n+2}}.
\label{ctau}
\end{equation}
Moreover, loops with virtual graviton exchanges can generate even
6-dimensional four-fermion interactions similar to the CI in
Eq.~(\ref{CI}). One example is the axial-axial operator:
\begin{equation}
{\cal L}_\Upsilon=\frac{1}{2}\, c_\Upsilon \left(\sum_f{\bar
f}\gamma_\mu\gamma_5 f\right)\hskip 3pt \left(\sum_f{\bar
f}\gamma^\mu\gamma_5 f\right),
\label{dim-6}
\end{equation}
with
\begin{equation}
c_\Upsilon=\frac{\pi^{n-2}}{16\Gamma^2(n/2)}\hskip 3pt
\frac{\Lambda_{\rm NDA}^{2+2n}}{M_D^{4+2n}}.
\label{cupsilon}
\end{equation}
\par
The current experimental lower bounds on the mass scales in Eqs.~(\ref{CI})
and (\ref{dim-8}), that parametrize the strength of the corresponding contactlike interactions, can be summarized qualitatively as follows~\cite{Yao:2006px}:
$\Lambda_H>1.3\, {\rm TeV}$;
$\Lambda_{\alpha\beta}> 10-15\, {\rm TeV}$ [95\% C.L.].

\section{Discovery and identification of the ADD scenario}

Clearly, constraints on $\Lambda_{\alpha\beta}$ and $\Lambda_H$
are determined by the deviations of the observables, ${\cal O}$,
from the SM expectations. We choose as basic observables the
longitudinally polarized differential cross sections, ${\cal
O}={\rm d}\sigma/{\rm d}\cos\theta$, for the fermionic processes
$e^+e^-\to{\bar f}f$ at ILC ($f$ is limited to $e,\mu,\tau,c,b$).
Obviously, the theoretical expressions of the cross sections
including the novel physics (NP), to be compared to the data, are
given by ${\rm d}\sigma\propto\vert{\rm SM}+{\rm
NP}(\Lambda)\vert^2$, where $\Lambda$ generically denotes
$\Lambda_{\alpha\beta}$ or $\Lambda_H$. It has been strongly
emphasized~\cite{Moortgat-Pick:2005cw} that electron and positron
beams polarization plays a crucial r\^ole in enhancing the
sensitivity to the NP interactions and, indeed, this option is
very seriously considered for the planned ILC.
\par
The comparison between ``theoretical'' relative deviations, $\Delta{\cal O}$,
and corresponding foreseen experimental relative uncertainties,
$\delta{\cal O}$, can be performed by a simple $\chi^2$ procedure combining
the initial polarization configurations and the binning of the angular range
for the measured reactions~\cite{Pankov:2005kd,Pankov:2006qi}:
\begin{equation}
\Delta{\cal O}=\frac{{\cal O}(\rm SM+NP)-{\cal O}(\rm
SM)}{{\cal O} (\rm SM)},\qquad \chi^2({\cal O})=\sum_{\{P^-,\ P^+\}}
\sum_{\rm bins}\left(\frac{\Delta({\cal O})^{\rm bin}}
{\delta{\cal O}^{\rm bin}}\right)^2.
\label{reldev}
\end{equation}
The $\chi^2$ in Eq.~(\ref{reldev}) will be a function of the mass
scale $\Lambda$ relevant to the contactlike interaction under
consideration. The expected {\it discovery} reach on an individual
interaction, i.e., the maximum value of the corresponding mass
scale $\Lambda$ for which a deviation caused by the interaction
itself could be observed, can
be assessed by assuming a situation where no deviation is observed
and imposing, for 95\% C.L., the constraint $\chi^2\le 3.84$.
Basically, this is the way the current limits above have been
obtained.
%%%%%%%%%%%%%%%%%%%%%%%%%%%%%%%%%%%%%%%%%%%%%%%%%%%%%%%%%%%%%%%%%%%%%%%%%%%%
%\par
%Current experimental limits, from LEP2 and Tevatron, are in the
%range $\Lambda_H>1.1-1.3$ TeV
%\cite{Ask:2004dv,Unel:2004fn,Landsberg:2004mj,Abazov:2005tk}. For
%the ${\rm TeV}^{-1}$-scale extra dimension scenario the limit,
%mostly determined by LEP data, is $M_C>6.8\hskip 3pt{\rm TeV}$
%\cite{Cheung:2004ab}.
%*************************************************
\par
In Table~\ref{table:1}, we give examples of discovery reaches
expected for an ILC with the ``reference'' parameters: $\sqrt
s=0.5$ TeV; time-integrated luminosity ${\cal L}_{\rm int}=500\,
{\rm fb}^{-1}$, and electron and positron longitudinal
polarizations $\vert P^-\vert=0.8$, $\vert P^+\vert=0.3$. While
these luminosity and beams polarization seem guaranteed at the
initial stage of ILC, ${\cal L}_{\rm int}=1000\, {\rm fb}^{-1}$
and $\vert P^+\vert$ of the order of 0.6 may be considered,
eventually, for later runs of the machine. To obtain the results
in Table~\ref{table:1}, binning of the angular range by
$\Delta\cos\theta=0.2$ intervals has been used in (\ref{reldev}),
and the statistical uncertainties have been evaluated by the final
fermions reconstruction efficiencies: 100\% for electrons, 95\%
for $\mu$ and $\tau$, 35\% and 60\% for $c$ and $b$ quarks,
respectively. The dominant systematic uncertainties are found to
originate from polarizations and luminosity, on which we have
assumed the accuracies 0.1\% and 0.5\%, respectively. Earlier
determinations, demonstrating the fundamental r\^ole of beams
polarization for the discovery reaches on CI interactions, can be
found, e.g., in Ref.~\cite{Riemann:2001bb}. The Table~1 shows the
high sensitivity to $\Lambda_{\alpha\beta}$ allowed by
polarization, and that Bhabha scattering is the process most
sensitive to $\Lambda_H$.
%------------------------------------------------------
\begin{table}[!htb]
\begin{center}
\begin{tabular}{|l|ll|ll|ll|ll|}
\hline
 & \multicolumn{8}{c|}{Processes} \\
\multicolumn{1}{|c|}{Model} & \multicolumn{2}{c|}{$e^{+}e^{-} \to e^{+}e^{-}$} & \multicolumn{2}{c|}{$e^{+}e^{-} \to l^{+}l^{-}$} & \multicolumn{2}{c|}{$e^{+}e^{-} \to \bar{b}b$} & \multicolumn{2}{c|}{$e^{+}e^{-} \to \bar{c}c$} \\
\hline
$ \Lambda_{H} $ & \multicolumn{1}{r}{5.3; } & 5.5 & \multicolumn{1}{r}{3.7; } & 3.8 & \multicolumn{1}{r}{3.7; } & 4.0 & \multicolumn{1}{r}{3.7; } & 3.8 \\
$\Lambda_{VV}^{ef}$ & \multicolumn{1}{r}{128.3; } & 136.7 & \multicolumn{1}{r}{136.4; } & 144.2 & \multicolumn{1}{r}{115.8; } & 137.4 & \multicolumn{1}{r}{128.3; } & 136.7 \\
$\Lambda_{AA}^{ef}$ & \multicolumn{1}{r}{76.1; } & 90.3 & \multicolumn{1}{r}{122.4; } & 129.5 & \multicolumn{1}{r}{116.7; } & 139.5 & \multicolumn{1}{r}{116.9; } & 124.8 \\
$\Lambda_{LL}^{ef}$ & \multicolumn{1}{r}{66.2; } & 82.7 & \multicolumn{1}{r}{81.9; } & 98.6 & \multicolumn{1}{r}{96.9; } & 105.7 & \multicolumn{1}{r}{84.1; } & 96.6 \\
$\Lambda_{RR}^{ef}$ & \multicolumn{1}{r}{64.0; } & 81.5 & \multicolumn{1}{r}{78.4; } & 97.7 & \multicolumn{1}{r}{64.4; } & 98.0 & \multicolumn{1}{r}{71.5; } & 95.3 \\
$\Lambda_{LR}^{ef}$ & \multicolumn{1}{r}{94.9; } & 100.1 & \multicolumn{1}{r}{74.1; } & 90.2 & \multicolumn{1}{r}{76.0; } & 95.9 & \multicolumn{1}{r}{54.5; } & 79.0 \\
$\Lambda_{RL}^{ef}$ & \multicolumn{2}{c|}{$\Lambda_{RL}^{ee} = \Lambda_{LR}^{ee}$} & \multicolumn{1}{r}{74.0; } & 90.6 & \multicolumn{1}{r}{70.9; } & 85.5 & \multicolumn{1}{r}{78.2; } & 86.5 \\
$M_{C}$ & \multicolumn{1}{r}{20.5; } & 22.1 & \multicolumn{1}{r}{30.7; } & 32.5 & \multicolumn{1}{r}{9.7; } & 14.9 & \multicolumn{1}{r}{15.8; } & 17.3 \\
\hline
\end{tabular}
\end{center}
 \caption{\label{table:1}   95\% C.L. discovery reaches (in TeV).
Left and right entries in each column refer to the polarizations
$(\vert P^-\vert,\vert P^+\vert$)=(0,0) and (0.8,0.3),
respectively. }
\end{table}
%------------------------------------------------------
\par
In principle, different
interactions may cause similar deviations in (\ref{reldev}), and
one would need to identify, among the various contact
interactions, the origin of the deviations, were they observed. In this regard, the {\it identification reach} on a given contact effective
interaction can be defined as the maximum value of the characteristic mass
scale $\Lambda$ for which the considered interaction not only can cause
observable deviations from the SM, but can also be discriminated as the
source of the observed deviations against the other contact interactions
for all values of {\it their} respective $\Lambda$s.
\par
Earlier attempts to estimate the {\it identification reaches} on ADD and CI models in
high energy $e^+e^-$ reactions have been presented in Ref.~\cite{Pasztor:2001hc}.
We here continue with the $\chi^2$ analysis outlined above \cite{Pankov:2005kd,Pankov:2006qi}.
\par
To make an illustrative example, we assume that the
ADD model (\ref{dim-8}) is found to be consistent with observed deviations.
To assess the level at which this scenario can be distinguished from each of the CI
models of Eq.~(\ref{CI}), one can consider the ``distances'' in the
($\Lambda_H,\Lambda_{\alpha\beta}$) two-dimensional planes:
\begin{equation}
{\tilde\Delta} ({\cal O})= \frac{{\cal O}({\rm CI})-{\cal O}({\rm
ADD})}{{\cal O}({\rm ADD})}, \qquad\quad {\tilde\chi}^2({\cal O})=
\sum_{\{P^-,\ P^+\}}\sum_{\rm bins}\left (\frac{{\tilde\Delta}({\cal
O})^{\rm bin}} {{\tilde\delta}{\cal O}^{\rm bin}}\right)^2.
\label{chitilde}
\end{equation}

\begin{wrapfigure}{r}{0.5\columnwidth}
\centerline{\includegraphics[width=0.5\columnwidth]{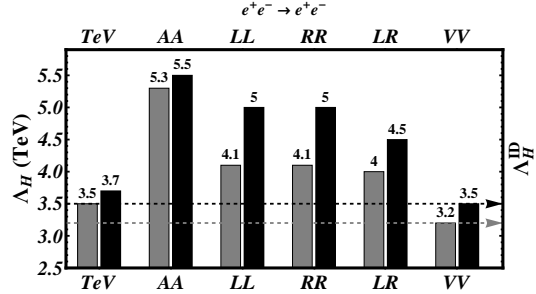}}
\vspace*{-0.6cm} \caption{Exclusion and identification reaches on
$\Lambda_H$ at 95\% C.L. obtained from
%the
Bhabha scattering.
%at $\sqrt{s}=0.5$ TeV, ${\cal L}_{\rm int}=500$ fb$^{-1}$ for
%unpolarized beams (gray histograms) and both beams polarized
%(black histograms) with $(\vert P^-\vert,\vert
%P^+\vert$)=(0.8,0.3) are illustrated.
} \label{Fig:Histogr} \vspace*{-0.1cm}
\end{wrapfigure}

In Eq.~(\ref{chitilde}), symbols are analogous to
Eq.~(\ref{reldev}), except that the statistical component of the
uncertainty ${\tilde\delta}\cal O$ is now referred to the ADD
model prediction. For each pair of $\alpha,\beta$ subscripts, we
can find {\it confusion} regions in the above mentioned planes,
where models cannot be distinguished from each other at the 95\%
C.L., by imposing the conditions
${\tilde\chi}^2(\Lambda_H,\Lambda_{\alpha\beta})\le 3.84$ for the
pairs $\alpha\beta= {\rm LL,RR,RL,LR}$. Each confusion region is
enclosed by a contour that shows a minimum value of $\Lambda_H$,
$\Lambda_H^{(\alpha\beta)}$, below which there is no confusion,
namely, the ``$\alpha\beta$'' CI model can be {\it excluded} as
the source of the observed deviations for all values of
$\Lambda_{\alpha\beta}$. Clearly, the smallest of the
$\Lambda_H^{(\alpha\beta)}$ determines the expected identification
reach on the ADD model (\ref{dim-8})~\cite{Pankov:2005kd}. This is
exemplified in Figure~\ref{Fig:Histogr}, that refers to an ILC
with $\sqrt s=0.5\, {\rm TeV}$, ${\cal L}_{\rm int}=500\, {\rm
fb}^{-1}$ unpolarized (grey bars) and with polarized beams with
$\vert P^-\vert=0.8$, $\vert P^+\vert=0.3$ (black bars). The
Figure indicates $\Lambda_H^{\rm ID}=$3.2 TeV (3.5 TeV)  as the
expected identification reach on (\ref{dim-8}) for unpolarized
(polarized) beams. The beams polarization, when combined as in
(\ref{chitilde}), play a r\^ole in substantially restricting the
confusion regions. This is even more evident by repeating the same
procedure for the identification reaches on the CI
couplings~\cite{Pankov:2005kd}.

\section{Model-independent identification of the ADD scenario}

\begin{figure}[!htb]
\vspace*{-1.0cm} \centerline{ \hspace*{-0.3cm}
\includegraphics[width=6.4cm,angle=0]{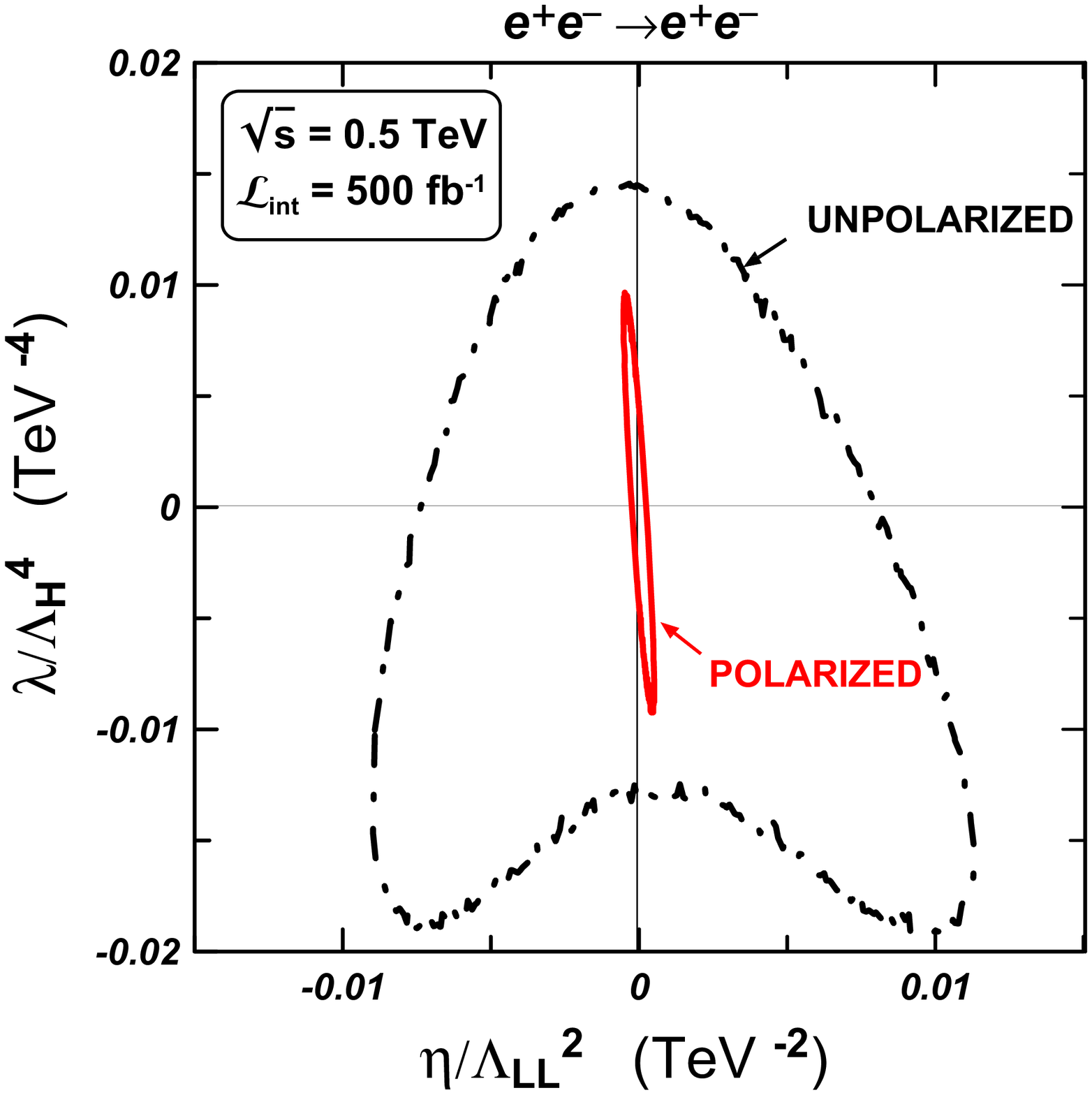}
%\hspace*{0.25cm}
\includegraphics[width=6.4cm,angle=0]{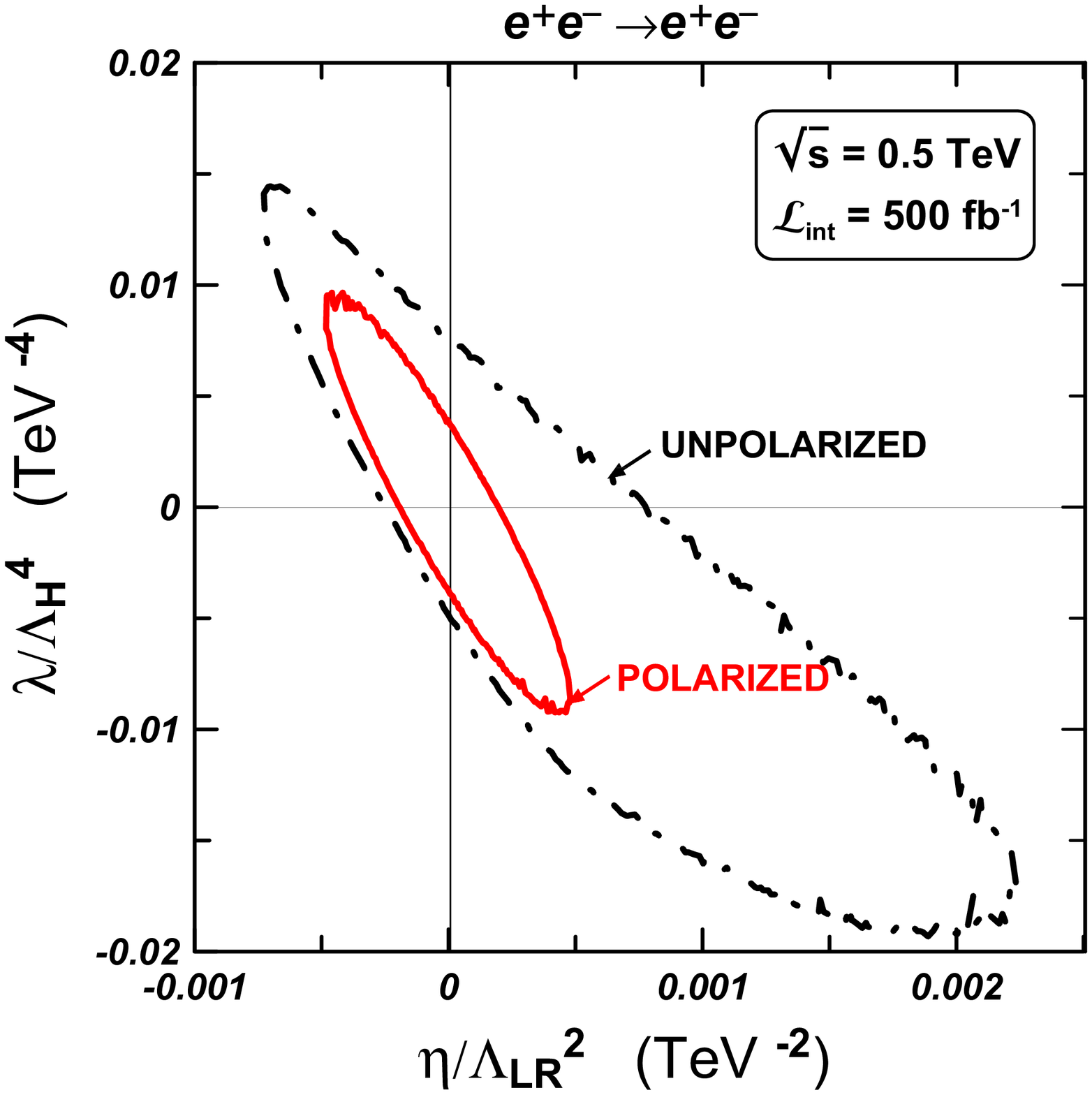}}
\vspace*{-1.8cm} \caption{\label{fig:1} Two-dimensional projection
of the 95\% C.L. confusion region onto the planes $(\eta_{\rm
LL}/\Lambda_{\rm LL}^2$, $\lambda/\Lambda_H^4)$
%(left panel)
and
$(\eta_{\rm LR}/\Lambda_{\rm LR}^2$ , $\lambda/\Lambda_H^4)$
%(right panel) obtained
from Bhabha scattering
%with unpolarized beams (dot-dashed curve) and with both beams polarized with
%$(\vert P^-\vert,\vert P^+\vert$)=(0.8,0.3) (solid curve)
. } \label{Fig:2}
\end{figure}

In the previous section we compared {\it pairs} of individual
contactlike interactions, (\ref{CI}) and (\ref{dim-8}). More
generally, we can consider the possibility that, for a given final
fermion flavour $f$, the CI interaction can be a linear
combination of {\it all} the individual interactions in
Eq.~(\ref{CI}) with free, simultaneously non vanishing,
independent coupling constants
$\eta_{\alpha\beta}/\Lambda^2_{\alpha\beta}$. In this case, the
corresponding identification reach on $\Lambda_H$ would be defined
as {\it model-independent}. The observables and their deviations
in Ref.~(\ref{chitilde}) now simultaneously depend on {\it all}
mass scales $\Lambda_{\alpha\beta}$ {\it and} $\Lambda_H$ as
${\cal O}({\rm CI})={\cal O}(\Lambda_{\rm LL},\Lambda_{\rm
RR},\Lambda_{\rm RL}, \Lambda_{\rm LR})$. The {\it confusion}
region in the multi-parameter space
($\Lambda_H,\Lambda_{\alpha\beta}$) with $\alpha,\beta={\rm L,R}$,
where the general CI model can mimic the ADD model and therefore
cannot be discriminated, is determined by the condition
${\bar\chi}^2\le{\bar\chi}^2_{\rm crit}$. Here, for 95\% C.L.,
${\bar\chi}^2_{\rm crit}=9.49$ for the annihilation channels
$f=\mu,\tau,c,b$ and ${\bar\chi}^2_{\rm crit}=7.82$ for Bhabha
scattering ($f=e$), where the $\rm LR$ and $\rm RL$ couplings are
equal. As an illustration, we show in Figure~\ref{Fig:2} examples
of the two-dimensional projections of the four-dimensional surface
enclosing the 95\% C.L. confusion region, onto the planes
($\eta_{\rm LL}/\Lambda_{\rm LL}^2,\lambda/\Lambda_H^4$) and
($\eta_{\rm LR}/\Lambda_{\rm LR}^2,\lambda/\Lambda_H^4$) for the
cases of unpolarized beams (dashed curves) and both beams
polarized with $(\vert P^-\vert,\vert P^+\vert)=(0.8,0.3)$ (solid
lines). As one can see, the r\^ole of polarization in restricting
the confusion region is dramatic.
%%%%%%%%%%%%%%%%%%%%%%%%%%%%%%%%%%%%%%%%%%%%%%%%%%%%%%%%%%%%%%%%%%%%%%%%%%
%%%%%%%%%%%%%%%%%%%%%%%%%%%%%%%%%%%%%%%%%%%%%%%%%%%%%%%%%%%%%%%%%%%%%%%%%%
\par
As indicated by Figure~\ref{Fig:2}, the contour of the confusion
region identifies a minimal value of $\Lambda_H$ for
which the CI scenario can be excluded as the source of the
deviations, and we take that value as the expected
model-independent identification reach on the ADD scenario
(\ref{dim-8})~\cite{Pankov:2006qi}. The numerical results for such
{\it model-independent identification reach}
$\Lambda_H^{\rm ID}$ at the ILC, with parameters exposed in the
caption, are shown in Table~\ref{table:IDR}.
%%%%%%%%%%%%%%%%%%%%%%%%%%%%%%%%%%%%%%%%%%%%%%%%%%%%%%%%%%%%%%%%%%%%%%%%%%%
\begin{table}[!htb]
\begin{center}
\begin{tabular}{|l|c|c|}
\hline \raisebox{-1.50ex}[0cm][0cm]{$\Lambda _{H}$ (TeV)}&
\multicolumn{2}{c|}{Process} \\
 & $e^{+}e^{-} \to e^{+}e^{-}$ & combined $e^{+}e^{-} \to \bar{f}f$ \\
\hline %\\[-0.3cm]
% \\[-0.3cm]
 ${\cal L}_{\rm int} = 500 fb^{-1}$ & 3.2& 4.8
\\ \hline %\\[-0.3cm]
${\cal L}_{\rm int} = 1000 fb^{-1}$ & 3.9& 5.2
\\ \hline %\\[-0.3cm]
\end{tabular}
\end{center}
\vspace*{.2cm} \caption{95\% C.L. model-independent identification
reach on $\Lambda_{H}$ obtained from Bhabha scattering and
combination of all final fermions ($f=e,\mu,\tau,c,b$) at
 $\sqrt{s} = 0.5$ TeV, ${\cal L}_{\rm int} = 500 fb^{-1}$, ($|P^-|$,$|P^+|$)=(0.8, 0.3) and
${\cal L}_{\rm int} = 1000 fb^{-1}$, ($|P^-|$,$|P^+|$)=(0.8, 0.6),
respectively.} \label{table:IDR}
\end{table}
%%%%%%%%%%%%%%%%%%%%%%%%%%%%%%%%%%%%%%%%%%%%%%%%%%%%%%%%%%%%%%%%%%%%%%%%%%%

\par
\begin{wrapfigure}{r}{0.5\columnwidth}
\vspace*{-0.5cm}
%\hspace*{1.0cm}
\centerline{\includegraphics[width=0.5\columnwidth]{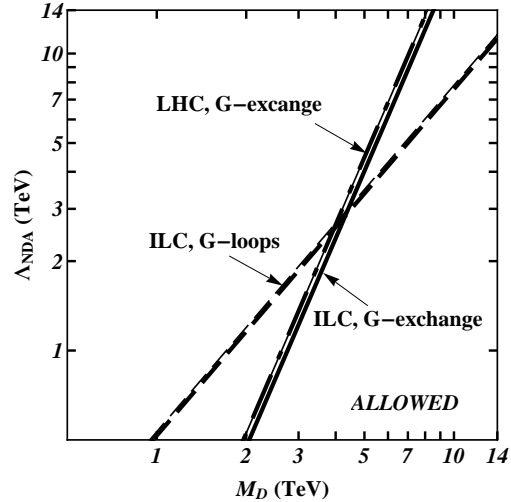}}
%\vspace*{-8.0cm}
\caption{95\% C.L. identification reaches obtained at the
polarized ILC(0.5 TeV) and LHC. } \label{Fig:3}
%\vspace*{-1.1cm}
\end{wrapfigure}

Using Eq.~(\ref{ctau}), we can turn the identification reach on
$\Lambda_H$ obtained above, into {\it allowed} and {\it excluded}
regions in the two-dimensional ($M_D,\Lambda_{\rm NDA}$) plane at
95\% C.L. An example, with $n=5$ and using the constraints
expected from combined fermionic processes $e,\mu,\tau,c,b$, is
shown in Figure~\ref{Fig:3} by the lines ``ILC, G-exchange'' for
the two options: ${\cal L}_{\rm int}=500\, {\rm fb}^{-1}$, $\vert
P^-\vert=0.8$, $\vert P^+\vert=0.3$ (thin solid curve) and ${\cal
L}_{\rm int}=1000\, {\rm fb}^{-1}$, $\vert P^-\vert=0.8$, $\vert
P^+\vert=0.6$ (thick solid curve).
\par
Analogously, one can derive the identification reach on the
coupling constant $c_{\Upsilon}$ in Eq.~(\ref{dim-6}), and then
the corresponding 95\% constraints in the ($M_D,\Lambda_{\rm
NDA}$) plane {\it via} Eq.~(\ref{cupsilon}). The results, under
the same conditions, are shown by the dashed lines ``ILC,
G-loops'' in Figure~\ref{Fig:3}. More details can be found in
Ref.~\cite{Pankov:2006qi}.
\par
It should be interesting to compare our results on the $M_D$ {\it
v.s.} $\Lambda_{\rm NDA}$ allowed regions with the expectations
from lepton-pair production $p+p\to l^+l^-+X$ ($l=e,\mu$) at the
LHC (DY). We qualitatively assume that the same value of $\Lambda$
enters into the different quark, antiquark and gluon subprocesses
relevant to DY. Also, we attempt to assess the discrimination of
deviations from the SM predictions caused by dimension-8
tree-level exchanges, Eqs.~(\ref{dim-8}) and (\ref{ctau}), from
those due to the dimension-6 AA four-fermion interaction,
Eqs.~(\ref{dim-6}) and (\ref{cupsilon}).
%%%%%%%%%%%%%%%%%%%%%%%%%%%%%%%%%%%%%%%%%%%%%%%%%%%%%%%%%%%%%%%%%%%%
%%%%%%%%%%%%%%%%%%%%%%%%%%%%%%%%%%%%%%%%%%%%%%%%%%%%%%%%%%%%%%%%%%%%%%
To this purpose, we utilize for the DY at the LHC the integrated
angular ``center-edge'' asymmetry proposed
in~\cite{Dvergsnes:2004tw}. This observable has the property of
being sensitive only to deviations from Eq.~(\ref{dim-8}), but
``transparent'' to those from both Eq.~(\ref{CI}) and
Eq.~(\ref{dim-6}).   The identification reach obtained from DY at
the LHC with ${\cal L}_{\rm int}=100\, {\rm fb}^{-1}$ (thick
dot-dashed curve) is shown in Figure~3.
\par
As Figure~3 shows, the limits on the tree-level graviton exchange
parametrized by Eq.~(\ref{dim-8}) and obtained from the LHC and
ILC are complementary rather than competitive. Moreover,
graviton-loop effects can dominate over tree-level exchange at
larger $M_D$. In this regime, the identification of the effective
operator $c_\Upsilon$ in fermion pair production at ILC provide
the most efficient probe of theories with extra dimensions. In
this case, the ILC(0.5 TeV) for chosen values of the luminosity
and beams polarization could be definitely superior to the LHC.

\section{Acknowledgments}
%AAP acknowledges the support of INFN and of MIUR (Italian Ministry
%of University and research). NP has been partially supported by
%funds of MIUR and of the University of Trieste.
This work is partially supported by the ICTP through the
OEA-Affiliated Centre-AC88.

% ****************************************************************************
% BIBLIOGRAPHY AREA
% ****************************************************************************

\begin{footnotesize}

% ----------------------------------------------------------------------------

\end{footnotesize}

\end{document}